# Population inversion and ultrafast terahertz nonlinearity of transient Dirac fermions in Cd$_3$As$_2$

Changqing Zhu,[1] Patrick Pilch,[1] Renato M. A. Dantas,[2,3,4,*] Anneke Reinold,[1] Yunkun Yang,[5,6,7] Faxian Xiu,[5,6] Amilcar Bedoya-Pinto,[8,9] Stuart S. P. Parkin,[8] Roderich Moessner,[2] Zhe Wang[1,*]

[1]*Department of Physics, TU Dortmund University, Dortmund, Germany*
[2]*Max Planck Institute for the Physics of Complex Systems, Dresden, Germany*
[3]*Department of Physics, University of Basel, Basel, Switzerland*
[4]*Center of Physics, University of Minho, Braga, Portugal*
[5]*State Key Laboratory of Surface Physics and Department of Physics, Fudan University, Shanghai, China*
[6]*Shanghai Research Center for Quantum Sciences, Shanghai, China*
[7]*Beijing Academy of Quantum Information Sciences, Beijing, China*
[8]*Max Planck Institute for Microstructure Physics, Halle (Saale), Germany*
[9]*Institute of Molecular Science, University of Valencia, Paterna, Spain*
(Dated: January 15, 2024)

Harmonic generation provides an efficient tool for the study of ultrafast nonlinear dynamics. We report on time-resolved optical-pump terahertz-harmonic-generation spectroscopic investigation of ultrafast nonlinearity in a prototypical three-dimensional Dirac semimetal Cd$_3$As$_2$. A transient population inversion characterized by excessive nonthermal Dirac electrons and holes is found to be very sensitive and responsive to a periodic terahertz drive, leading to very efficient terahertz third-harmonic generation. Based on the Boltzmann transport theory, we analyze the terahertz field-driven kinetics of the transient Dirac fermions that is responsible for the observed strong terahertz nonlinearity.

Transient states of matter far from thermal equilibrium exhibit a rich variety of novel physical phenomena[1,2,3,4]. Investigation of these phenomena not only enriches our understanding of the nonthermal states, but also offers unconventional means towards ultrafast control of quantum materials[2,4]. In particular, the nonlinear optical response of Dirac or Weyl semimetals is a very compelling subject (see e.g. Ref. [5,6,7,8,9,10,11,12,13,14,15,16,17,18,19,20,21,22,23,24,25,26,27,28,29,30]). For example, transition metal monopnictide Weyl semimetals with inversion-symmetry breaking exhibit giant second-order nonlinear polarizability in the near infrared to visible frequencies[7]; Due to Berry curvature dipole, second-order topological response leads to nonlinear Hall effects[5,17,21] or nonlinear Kerr rotation[6], which is applicable for photodetection[17]; Novel physical phenomena emerge also from third- or higher-order nonlinearity of Dirac or Weyl semimetals, such as the circular photogalvanic effect[11] or high-harmonic generation[8,9,10,12,13,14,18,19]. The studies of these phenomena allow to address physical problems, ranging from dynamical Bloch processes[8,10] and topological effects[18,19,23] to strong electronic correlations[16,27], and also provide novel possibilities for applications[8,10,11,15,20,28,29].

Recently in the prototypical three-dimensional (3D) Dirac semimetal Cd$_3$As$_2$ [31,32,33,34], a peculiar far-from-equilibrium state of population inversion was revealed by time- and angle-resolved photoemission spectroscopy (TrARPES)[35]. As the interband relaxation is much slower than for 2D Dirac fermions[36], the observed population inversion in Cd$_3$As$_2$ is long-lived with a lifetime of several picoseconds (ps)[35] in contrast to 130 femtoseconds (fs) in the 2D system[36]. Therefore, this 3D Dirac semimetal offers an opportunity particularly suitable for the investigation of transient nonthermal states in the terahertz (1 THz ~ 1 ps) frequencies, whereas faster dynamical processes were often studied in mid/near infrared, visible, or (extreme) ultraviolet[1,3,4]. In this work, we investigate the THz dynamics far from equilibrium and nonlinear responses of photoexcited transient Dirac fermions corresponding to the population inversion in Cd$_3$As$_2$. Our study reveals that the transient Dirac fermions of the population-inversion state react rapidly and intensely to a periodic THz drive, leading to strong THz third-harmonic generation (THG).

High-quality thin films of Cd$_3$As$_2$ are grown by molecular-beam epitaxy, see Ref. [37] for more details. For our optical experiment, the sample surface is parallel to the crystallographic (112) plane with a thickness of about 60 nm. Time-resolved nonlinear spectroscopic experiments are carried out based on a femtosecond laser system for wavelength of 800 nm, pulse duration of 100 fs, and repetition rate of 1 kHz. Broadband THz radiation is generated through a tilted pulse-front scheme utilizing a LiNbO$_3$ crystal (see e.g. Refs. [38,39,40]), while narrowband THz radiation is obtained by using a bandpass filter. The THz electric field is characterized through electro-optical sampling at a ZnTe crystal[41].

Cd$_3$As$_2$ is a well-established 3D Dirac semimetal whose band structure in the vicinity of the Fermi level is characterized by one pair of Dirac cones without trivial parabolic bands[31,32,33,34]. The inversion symmetry of its crystal structure with a space group *P4$_2$/nmc* dictates a vanishing second- or higher even-order nonlinear optical response. Hence we focus on the study of third-order nonlinearity. To investigate the THz nonlinear response of the population-inversion state, we synchronize an 800 nm pump pulse to an intense multicycle THz drive pulse with a central frequency of $f = 0.35$ THz and a peak field of 120 kV/cm [see inset of Fig. 1(b) for an illustration]. According to previous reports[42,43], the initial interband photoexcitation of electrons and holes is followed by a fast relaxation towards the Dirac cones at the Fermi energy with a typical relaxation time $\leq 100$ fs, leading to the formation of population inversion at the Dirac cones since there are no other bands around the Fermi level[35].

With the 800 nm pump, we indeed observe a strong THz third-harmonic emission, whose temporal trace is presented in Fig. 1(a) for a pump fluence of 0.96 mJ/cm$^2$ at a time delay $t_\text{pd} = -2.3$ ps. The minus sign indicates that the 800 nm pump pulse arrives later than the maximum peak of the THz drive pulse, while the overall duration of the THz pulse is about 20 ps [see Fig. 2(c) for its waveform]. The spectrum from Fourier transformation of the time-domain THG signal is shown in Fig. 1(c). Besides a strong peak around *3f*, a weak peak corresponding to the fundamental frequency is still discernible. To quantify the nonlinear response of the population-inversion state, we measure the THz THG also without the 800 nm pump and display the result in Fig. 1(b) and 1(d) for the time and frequency domain, respectively. Without the pump, the peak electric field of the THG reduces to about 1/3, while the integrated THG intensity drops to only 10% of the value with the pump. This corresponds to a decrease of the third-order nonlinear susceptibility from $\chi^{(3)} = \frac{E_{(3f)}}{E_{(f)}^3} = 8.0 \times 10^{-13}$ to $2.5 \times 10^{-13}$ V⁻²cm². It is worth noting that broadband THz emission can be photoinduced in Dirac or Weyl semimetals due to photogalvanic effects (see e.g. Ref. [44]) or photothermoelectric effects (see e.g. Ref.[45]), which requires specific optical alignment and/or sample preparation. In our experiment we have avoided these conditions by an off-alignment of the pump pulse, therefore without the THz drive pulse the 800 nm pump pulse alone cannot lead to the observed strong third-harmonic radiation[41].

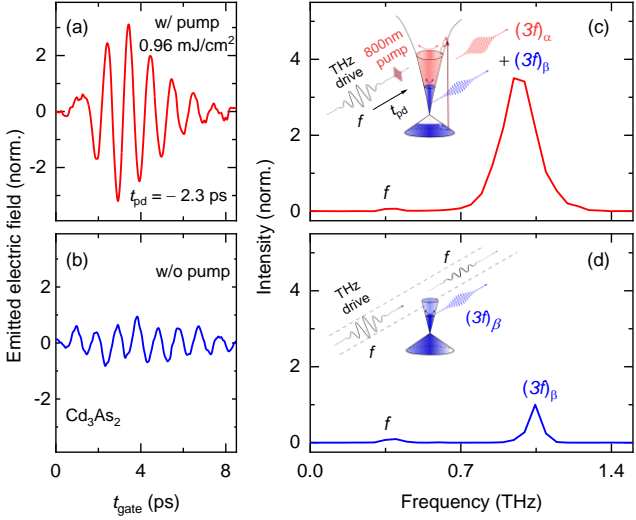
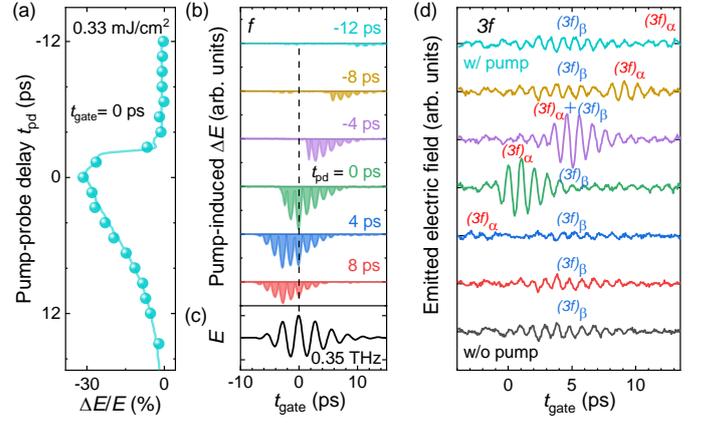

FIG. 1. (a) Emitted THz electric field from $Cd_3As_2$ with and (b) without an 800 nm optical pump of a fluence of 0.96 mJ/cm² at a time delay $t_{pd} = -2.3$ ps, which are recorded after a *3f*-bandpass filter. (c) and (d) Fourier transform spectra of the time-domain data in (a) and (b), respectively, exhibit third-harmonic generation. Insets in (c) and (d): Illustration of the THz THG measurements with and without 800 nm pump, respectively.

In comparison with the THG spectrum without pump, the peak position of the optically enhanced THG spectrum [cf. Fig. 1(c)(d)] is shifted slightly towards lower frequency by about 0.07 THz. As will be shown below, there are in fact two different THG components, as denoted by $(3f)_\alpha$ and $(3f)_\beta$, in the presence of the optical pump, where the $(3f)_\alpha$ signal appears due to the optical pump and characterizes the nonlinear response of the nonthermal states of the population inversion. Related to the optically induced $(3f)_\alpha$ signal the apparent slight frequency shift arises because the THz drive pulse has slightly more spectral weight on the lower-frequency side than on the higher-frequency side. In the presence of the optical excitation, the lower-frequency side of the THG gains stronger enhancement than the higher frequency side.

FIG. 2. (a) Optical pump-induced amplitude change of transmitted terahertz peak electric field $\Delta E/E$ (corresponding to $t_{gate} = 0$ ps) as a function of pump-probe delay $t_{pd}$. The optical pump fluence is 0.33 mJ/cm². (b) Pump-induced temporal amplitude depletion $\Delta E$ of the multicycle terahertz driving field for various positive and negative $t_{pd}$'s. The dashed line corresponds to the curve in (a). (c) Time trace of the multicycle terahertz driving field with a central frequency of $f = 0.35$ THz and peak field strength of 120 kV/cm. (d) Time traces of the third-harmonic radiation for the same $t_{pd}$'s in (b), which are recorded after a *3f*-bandpass filter. Two different THG traces are marked by $(3f)_\alpha$ and $(3f)_\beta$.

To understand the strong nonlinear response of the nonthermal states corresponding to the population inversion, we comprehensively characterize the linear and nonlinear responses of the optically excited nonequilibrium states. The results obtained through the linear- and nonlinear-response channels are presented in Fig. 2(a)(b) and Fig. 2(d), respectively, with varying the time delay $t_{pd}$ between the 800 nm pump and the THz pulse. We first measure the changes of the transmitted THz peak field $\Delta E/E$ induced by the 800 nm pump. As shown in Fig. 2(a), for a pump fluence of 0.33 mJ/cm² the transmitted THz field is reduced by up to 30%. This reflects a transient enrichment of electrons and holes due to the photoexcitation, which has been directly observed by TrARPES[35]. The relaxation process shown in Fig. 2(a) can be well approximated by a single exponential decay function with a characteristic relaxation time of 7.7 ps, in good agreement with previous reports[42,43]. According to the TrARPES study[35], this relaxation process corresponds to electron-hole recombination from the transient population-inversion state at the Dirac nodes.

For various time delays $t_{pd}$, the pump-induced amplitude changes of the overall waveform of the multicycle THz pulse are presented in Fig. 2(b), while the original THz waveform is given in Fig. 2(c). The pump-induced changes are always amplitude depletion of the terahertz electric field, meaning an enhanced absorption due to the increased charge-carrier density of the nonequilibrium states, which occurs only after the arrival of the pump pulse at the sample. For a large negative time delay, e.g. $t_{pd} = -12$ ps, which means that the 800 nm pump pulse arrives way behind the THz peak, only a negligible amplitude change is detected at the tail of the THz pulse. In contrast, a strong amplitude change is observed around $t_{pd} = 0$ ps. These results clearly show that by using the multicycle THz pulse to probe the nonequilibrium states, we can properly capture the relatively slow interband relaxation dynamics corresponding to the electron-hole recombination at the Dirac cones. While the initial fast relaxation governed by electron-electron interactions is typically ≤ 100 fs, the resolved lifetime of the transient population-inversion state is rather long[35], e.g. 7.7 ps for a pump fluence of 0.33 mJ/cm$^2$. Such a long lifetime is several times greater than the period of the 0.35 THz drive (about 2.86 ps), which is essential for the observed strong transient THz nonlinearity.

The results obtained through the channel of the THz third-order response are summarized in Fig. 2(d). In comparison with the situation without optical excitation, a strongly enhanced terahertz THG signal is observed at $t_{pd} = 0$ ps, similar to the data in Fig. 1. By shifting $t_{pd}$ towards the negative direction, the third-harmonic yield first remains almost invariant and then decreases continuously. A new and important dynamic feature appears when the 800 nm pump pulse is shifted away but not too far from the THz maximum peak-field position. At $t_{pd} = -8$ ps, one can clearly see two different THG traces which are well separated in the time domain. The one appearing earlier (from $t_{gate} = 0$ to ~7.5 ps) is less intense than the one observed later (after $t_{gate} = 7.5$ ps), which are denoted by $(3f)_\beta$ and $(3f)_\alpha$, respectively. This new feature is crucial for our understanding of the strong nonlinear response of the population inversion.

In contrast to the $(3f)_\beta$ signal, which appears at the same time-delay as the THG signal without the optical pump [see Fig. 2(d)], the $(3f)_\alpha$ signal shifts in the time domain with varying $t_{pd}$. Therefore, the $(3f)_\alpha$ signal reflects the nonlinear THz response of the photoexcited transient Dirac electrons and holes. When the optical pump pulse is well ahead (e.g. $t_{pd} = 8$ ps) or behind (e.g. $t_{pd} = -12$ ps) the THz drive pulse, the $(3f)_\alpha$ signal essentially vanishes, leaving only the $(3f)_\beta$ signal observed in the time traces. In contrast, when the two pulses overlap, the two THG signals are not distinguishable in the time domain and the overall THG is significantly enhanced. A more detailed measurement as a function of $t_{pd}$ shows that the optically induced $(3f)_\alpha$ signal becomes resolvable for $t_{pd} \lesssim 5.4$ ps (see Ref. [41]). It should be noted that this time-delay should not be understood as a critical value e.g. for a phase transition, but rather reflects a characteristic time scale within which the nonlinear response of the population-inversion state is experimentally resolvable. With varying polarization of the 800 nm or the THz pulse, we found no evident polarization dependence[41], which is clearly different from other effects with strong polarization dependence, such as photogalvanic[44] or photothermoelectric effects[45].

To further investigate the THz nonlinearity due to the optically excited nonthermal Dirac fermions, we monitor the evolution of the third-harmonic radiation by gradually increasing the pump fluence. In particular, we fix the time-delay at $t_{pd} = 2.7$ ps, where we can clearly see two temporally separated THG signals. As shown in Fig. 3(a), while at zero fluence only the $(3f)_\beta$ radiation is observed, a weak $(3f)_\alpha$ signal is discernible already at a very low pump fluence of 42 μJ/cm$^2$ which appears well before the $(3f)_\beta$ traces. With increasing fluence the intensity of the $(3f)_\alpha$ signal increases continuously, whereas the $(3f)_\beta$ signal is substantially suppressed at high fluences. The fluence dependence of the integrated intensity for the two THG traces is summarized in Fig. 3(d). These results clearly show that the two THG signals are not only of different origin but also appear to compete with each other.

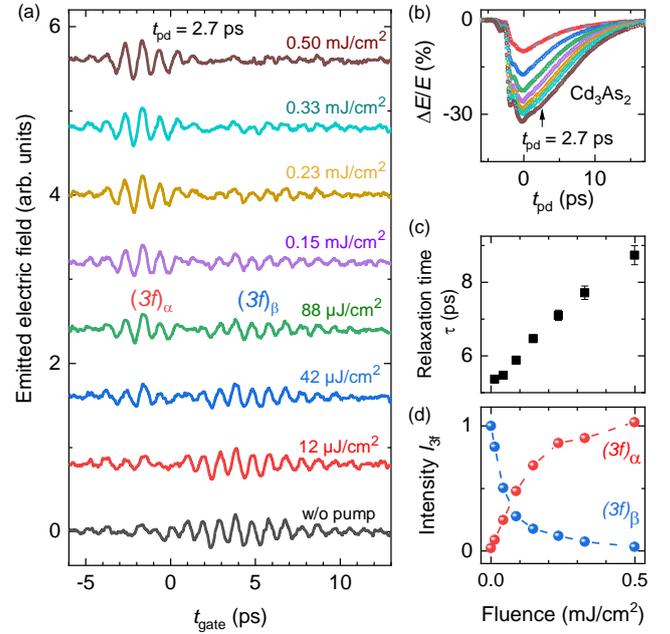

FIG. 3. (a) Temporal traces of THz THG for various 800 nm pump fluences at a time delay $t_{pd} = 2.7$ ps, which is indicated by the arrow in (b). In comparison with the trace without the pump, which is marked by $(3f)_\beta$, an additional third-harmonic trace is observed at an earlier time due to the optical pump, which is marked by $(3f)_\alpha$. (b) Pump-induced amplitude changes of the transmitted electric field of the THz drive pulse as a function of $t_{pd}$ for the corresponding pump fluencies in (a). The dashed line in (b) presents an exponential fit. (c) Relaxation time as a function of pump fluences as obtained by simulating the experimental data in (b) using an exponential function. (d) Intensity of the observed third-harmonic traces $(3f)_\alpha$ and $(3f)_\beta$ versus pump fluence.

We can understand the observed dynamical behavior by scrutinizing the pump induced THz absorption for the corresponding pump fluences [Fig. 3(b)]. With increasing fluence the absorption of the THz field is enhanced, and concomitantly the relaxation time increases from about 5 ps at 12 μJ/cm$^2$ to 9 ps at 0.50 mJ/cm$^2$ [Fig. 3(c)]. Since the slow relaxation processes correspond to electron-hole recombination at the Dirac cones[35], the increase of pump fluence not only leads to an enhanced density of the nonthermal Dirac fermions (electrons and holes), which is reflected by the increased THz absorption [Fig. 3(b)], but also gives rise to an increase in the lifetime of the population inversion [Fig. 3(c)]. These two effects are both in favor of yielding the $(3f)_\alpha$ radiation, which is generated nearly instantaneously after the optical pump, whereas the THG without the photoexcitation [i.e. the $(3f)_\beta$ traces] is rather retarded: An increased lifetime of the transient Dirac fermions of a few picoseconds allows the THz field to adequately drive the nonthermal Dirac electrons and holes before their recombination, leading to stronger nonlinear current and thereby highly enhanced third-harmonic generation; A higher density of excited Dirac fermions can carry a greater nonlinear current density. In contrast, the retarded $(3f)_\beta$ harmonic radiation is further reduced with increasing pump fluence, because at a higher pump fluence more transient Dirac fermions are created already at an earlier time, resulting in an increased absorption of the THz field in the linear response [see Fig. 2(b)] and an enhanced generation of the $(3f)_\alpha$ radiation.

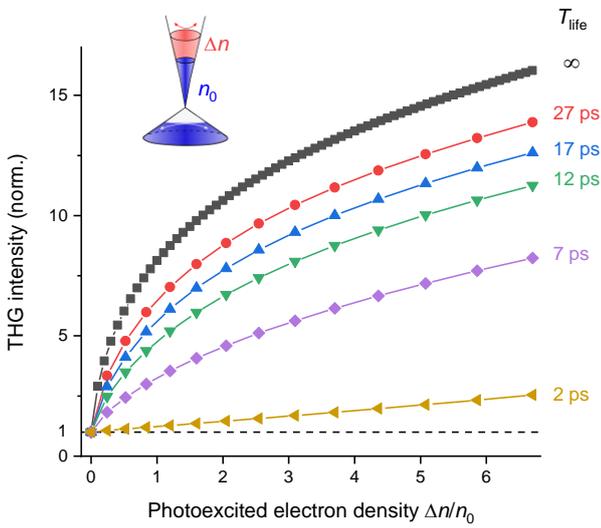

FIG. 4. Theoretical analysis of terahertz driven kinetics of population-inversion state. Normalized intensity of THz third-harmonic generation as a function of photoexcited electron density $\Delta n$ with respect to the electron density in the equilibrium state $n_0$. The lifetime $T_{\rm life}$ of the population inversion is varied from infinity to 2 ps. The time delay between the optical pump and the THz drive pulse is fixed to $t_{\rm pd} = 5.4$ ps. The inset illustrates a transient state of population inversion.

To gain a quantitative understanding of the observed transient nonlinearity, we analyze the THz field-driven kinetics of the Dirac fermions based on the Boltzmann transport theory[18,41]. To clarify the role of the long-lived population-inversion state in the observed third-harmonic emission, we study the dependence of the third-harmonic radiation on two crucial parameters – the lifetime of the population inversion $T_{\rm life}$ and the density of the photoexcited Dirac fermions $\Delta n / n_0$. Specifically, we consider the chemical potential to be time dependent, which decays exponentially and is given by

$$\mu^{(e)}(t) = \mu_0^{(e)} + \mu_0^{(e)} \left[ \left(1 + \frac{\Delta n}{n_0}\right)^{1/3} - 1 \right] e^{-\frac{t+t_{pd}}{T_{\rm life}}}$$

for electrons, and $\mu^{(h)}(t) = \frac{v_F^{(h)}}{v_F^{(e)}} \mu_0^{(e)} \left(\frac{\Delta n}{n_0}\right)^{1/3} e^{-\frac{t+t_{pd}}{T_{\rm life}}}$ for holes[41]. Here $n_0$ denotes the upper-band electron density without the photoexcitation (see Fig. 4 inset). We adopt the experimental parameters[32] of the Cd$_3$As$_2$ thin film for our simulation, i.e. the initial chemical potential $\mu_0^{(e)} = 0.2$ eV, Fermi velocity for electrons $v_F^{(e)} = 7.8 \times 10^5$ m/s, and holes $v_F^{(h)} = 1.3 \times 10^6$ m/s.

Figure 4 presents the theoretically obtained third-harmonic radiation as a function of photoexcited electron density for various lifetimes $T_{\rm life}$ of the population-inversion state. In this representation the time delay between the optical pump pulse and the THz drive pulse is fixed to $t_{\rm pd} = 5.4$ ps, which corresponds to the appearance of the THG in our experiment[41]. We note that the photoexcited hole density is the same as the photoexcited electron density, which is not explicitly mentioned in the representation. In our simulation, the parameter of chemical potential corresponds to the highest energy level one can reach by filling all electrons in the upper Dirac band.

The most efficient THG is observed for an infinite lifetime of the population-inversion state, i.e. $T_{\rm life} = \infty$, which corresponds to no electron-hole recombination. Starting from the initial equilibrium state, the third-harmonic emission is enhanced monotonically with increasing photoexcited charge carrier density. One can see from Fig. 4 that the efficiency of the THz third-harmonic yield drops evidently when the lifetime of the population-inversion state decreases. It is interesting to remark that even when $T_{\rm life}$ is slightly smaller than $t_{\rm pd}$, one can still observe weak but finite third-harmonic yield from the photoexcited states under the strong THz drive (see the $T_{\rm life} = 2$ ps curve in Fig. 4). This is because $T_{\rm life}$ is a characteristic value rather than a critical value. Within the time scale of $t_{\rm pd}$, a minor fraction of the photoexcited charges remains unrelaxed. However, when $t_{\rm pd}$ is significantly greater than $T_{\rm life}$, no enhancement of THG can be obtained which is consistent with the experimental observation.

To conclude, we have observed a strong THz nonlinear response of the optically excited transient Dirac electrons and holes in the 3D Dirac semimetal Cd$_3$As$_2$. The relevant transient far-from-equilibrium state is featured by a relatively long-lived population inversion of the electrons and holes at the Dirac cones with a picosecond characteristic lifetime. This population inversion under a strong terahertz drive yields very efficient third-harmonic radiation. The involved nonlinear kinetics can be understood based on the Boltzmann transport formalism. We anticipate that our findings will motivate further studies of the ultrafast nonlinearity not only for the Dirac or Weyl fermions in semimetals, but also for field driven charge and/or spin transport in semiconductors or magnetic heterostructures.

We thank Ahmed Ghalgaoui and Yang Zhang for helpful discussions. We acknowledge support by the European Research Council (ERC) under the Horizon 2020 research and innovation programme, grant agreement No. 950560 (DynaQuanta). F.X. was supported by the National Natural Science Foundation of China (52225207, 11934005, and 52350001), the Shanghai Pilot Program for Basic Research - FuDan University 21TQ1400100 (21TQ006) and the Shanghai Municipal Science and Technology Major Project (Grant No.2019SHZDZX01). A.B.-P. acknowledges support by the Generalitat Valenciana (CIDEGENT/2021/005). The work in Dresden is in part supported by the Deutsche Forschungsgemeinschaft (DFG) via SFB 1143 (project-id 247310070) and cluster of excellence ct.qmat (EXC 2147, project-id 390858490). R.M.A.D. acknowledges support from the European Commission through the project Graphene Driven Revolutions in ICT and Beyond (Ref. No. 881603, CORE 3).

C.Z. and P.P. contributed equally to this work.
\*renatodantas@fisica.uminho.pt (R.M.A.D.);
zhe.wang@tu-dortmund.de (Z.W.)